\documentstyle[floats,aps]{revtex}
\input{psfig}
\begin{document}
\def \beq{\begin{equation}}
\def \eeq{\end{equation}}
\def \beqarr{\begin{eqnarray}}
\def \eeqarr{\end{eqnarray}}
\def \be{\begin{equation}}
\def \ee{\end{equation}}
\def \bea{\begin{eqnarray}}
\def \eea{\end{eqnarray}}
\def \ta{{\tilde\alpha}}
\def \tg{{\tilde g}}     
\twocolumn[\hsize\textwidth\columnwidth\hsize\csname @twocolumnfalse\endcsname
\title{Non--linear Transport in the Quantized Hall Insulator}
\author{Efrat Shimshoni$^{1,2}$ and Assa Auerbach$^2$}
\address{$^1$ Department of Mathematics-Physics, Oranim--Haifa University,
Tivon 36006, Israel.}
\address{$^2$ Department of Physics, The Technion, Haifa 32000, Israel.}
\date{\today}
\maketitle
\begin{abstract}
We model the insulator neighboring the $1/k$ quantum Hall phase by a  random network of  puddles of filling fraction $1/k$.   The puddles are coupled by weak tunnel barriers. Using Kirchoff's laws we prove that the macroscopic Hall
resistivity is quantized at
$k h/ e^2$ and independent of magnetic field and  current bias -- in agreement with recent experimental 
observations.  In addition,  for $k>1$ this theory predicts non linear 
longitudinal response $V\sim I^\alpha$ at zero temperature, 
and $V/I\sim T^{1-{1\over \alpha}}$ at low bias. 
$\alpha$ is determined using  Renn and Arovas' theory for the 
single  junction response, and is related to  the Luttinger liquid spectra of
the edge states straddling  the typical tunnel barrier. The dependence of $V(I)$ on
the magnetic field  is related to the typical puddle size.
Deviations of $V(I)$ from a pure power are estimated using  a series/parallel approximation for the
two dimensional  random nonlinear resistor network. We check  the validity of this approximation by numerically solving for 
a finite square lattice network.
\end{abstract}
\vskip2pc]
\narrowtext
\section{Introduction}
\label{sec:intro}
The ``Hall insulator'' defines a peculiar insulating state, in which the
longitudinal resistivity $\rho_{xx}$ diverges in the limit of zero 
temperature and frequency, yet the Hall resistivity $\rho_{xy}$
remains {\it finite}. Such a behavior of $\rho_{xy}$ has been argued to be a
quite generic property of disordered single electron models\cite{Fuku,KLZlett,Joe},  provided
$\sigma_{xy}\sim\sigma^2_{xx}$ in the limit of vanishing conductivities
$\sigma_{xx}$ and $\sigma_{xy}$\cite{Ora}. That is to say,
\beq
\lim_{\sigma_{xx}\to 0}\rho_{xy}=\sigma_{yx}/(\sigma_{xx}^2+\sigma_{xy}^2) 
< \infty\; .
\label{rxydef}
\eeq
Experimentally,  
several groups have observed a Hall insulating behavior in strong 
magnetic fields --  both in three-dimensional samples \cite{Hop} and in
the quantum Hall (QH) regime \cite{Vladimir,SSS}. 
In the global phase diagram of Kivelson, Lee and Zhang \cite{KLZ} the 
entire insulating phase
surrounding the QH liquid phases  is predicted to exhibit
Hall insulating behavior with
$\rho_{xy}\sim B/nec$, as in a {\em classical} Hall conductor -- 
in agreement with the data of, e.g., Ref. \cite{Vladimir}. 
(Here $B$ is the perpendicular magnetic field and $n$ is the electron density).

Confusion has been compounded recently by measurement of the Hall voltage near the transition between a 
$1/3$--QH liquid and the insulator \cite{SSS} which indicates a different
behavior of $\rho_{xy}$ in the insulating phase: quite remarkably, it 
preserves the quantized value $3h/e^2$ over a finite range of the 
magnetic field (the parameter which drives the transition) {\it beyond} 
the critical point. Moreover, the Hall voltage
$V_H(I)$ is linear in the low current range where the longitudinal voltage
$V(I)$ exhibits an insulating--like non--linear dependence.
A deviation from the quantized Hall resistance, approaching a linear rise as a
function of $B$, is observed only deeper in the insulating phase. This 
persistence of the QH plateau can not be explained by means of transport models
based on hopping between strongly localized single--electron sites, as such
models do not pose any particular restriction on the {\it value} of the finite
Hall coefficient. 

In this paper we propose a transport model which reflects the prominent
phenomenology of the exotic insulating phase described above -- hereon dubbed
``a quantized Hall insulator'' (QHI).  It is clear that one needs to take into account 
both electron interactions, which are responsible for the  fractional QH effect, and the random potential.  This task is manageable in the limit of slow potential variations
with respect to the magnetic length. The  incompressibility of the electron liquid 
at magic densities,   creates
puddles of  QH liquid at these densities  in the shape of the equipotential contours.

Thus we extend Chalker and Coddington's network model \cite{CC}
from the integer to the fractional QH regime.
In place of the semiclassical single--electron orbits, the transport here is conducted through a random network of edge states surrounding 
the puddles of  $1/k$--QH liquid defined to be of density  
$n=B/(k\phi_0)$, where $\phi_0$ is a flux quantum. 
The edge states are connected to 
each other by tunnel barriers. As we show below, if the percolating network
of edge states belongs to a {\it single fraction} $1/k$, $\rho_{xy}$ of the 
entire network acquires the quantized value $k\hbar/e^2$ \cite{ruzin}.   
Since edge states
tunneling between  fractional QH liquid involves power--law density of states, the 
longitudinal current--voltage characteristic of the system,
$V(I)$ is generally non--linear and  $dV/dI\rightarrow \infty$ for $I\rightarrow 0$ in the
weak tunneling limit. Below, we 
calculate the Hall and longitudinal response of the edge states network, 
and relate its properties to the total carrier density, the magnetic field
and the potential fluctuations.

In the framework of the present paper we do not elaborate on the
justification of the puddles model; rather, we focus on its consequences
on transport properties, some of which are yet to be
confronted with experiment. It is worthwhile pointing out, though,
that the true ground state of certain realistic systems is quite
likely to be well imitated by such a model. In restricted regions
of the sample, where considerable variations in the disorder potential
occur over length scales much larger than the magnetic length, the
formation of puddles of incompressible liquid is energetically
favorable. Since transport inside such a puddle is dissipationless, a
current carrying path that crosses the entire sample is expected to be
dominated by channels that `hop'  between neighboring puddles 
at places of minimal separation.

In section \ref{sec:hall} we prove that the Hall resistance of a $1/k$ 
puddle--network is quantized at $\rho_{xy}=k\hbar/e^2$. In section 
\ref{sec:long} we calculate the non--linear longitudinal response $ V(I)$ 
of the network. Deviations from a pure power--law behavior are estimated in
Appendix A; the parameters of the model are related to the magnetic
field and potential fluctuations (using the theory of Renn and Arovas 
\cite{RA} for single QH tunnel junctions) in Appendix B. In section  
\ref{sec:summary} we summarize our main results and point out some
open questions and suggestions for experimental tests of our model.

\section{The Hall Resistance of a Puddle--Network}
\label{sec:hall}
\begin{figure}[htb]
\centerline{\psfig{figure=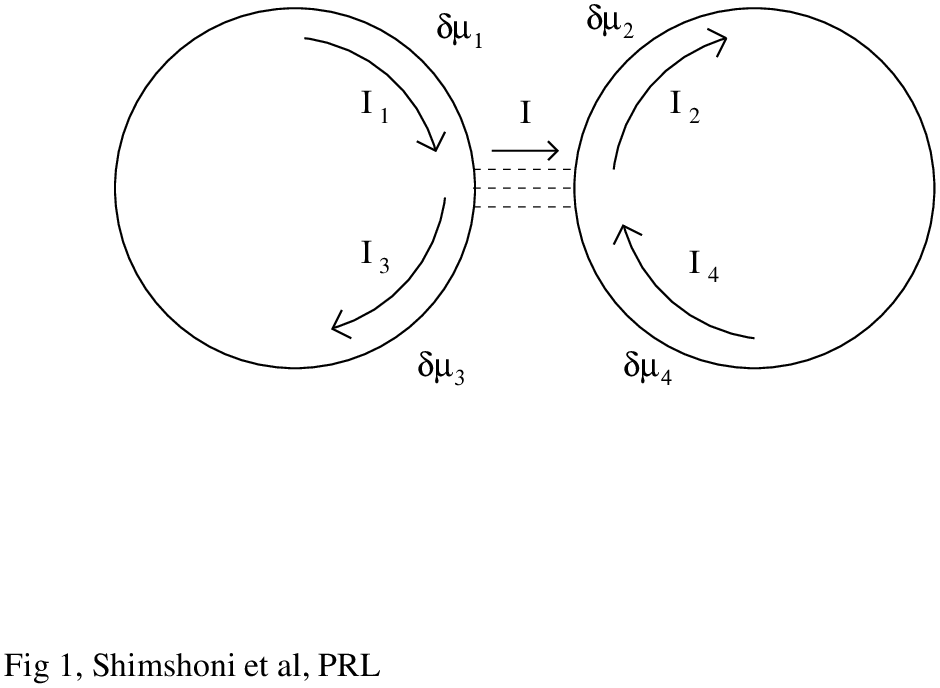,width=3in}}
\caption[]
{
\label{fig:fig1}
A single junction between puddles.
}
\end{figure}

Consider a random two-dimensional network, combined of the basic elements 
schematically depicted in Fig. \ref{fig:fig1}.
Circles denote the ``puddles''; each couple of puddles is separated by
a tunnel junction which involves four edge currents $I_1,..I_4$.
By current conservation the tunnelling current  $I$ is given by 
\be
I  = I_1-I_3=I_2-I_4
\label{Ii}
\ee
The macroscopic theory of a Hall liquid in a confining potential yields a fundamental  relation between the  
excess chemical potentials at the edges 
$\delta\mu_i$ and the edge currents\cite{Beenakker}:
\be
\delta\mu_i = \mbox{sgn}(B) {h\over e^2} k I_i
\label{Vi}
\ee
where   $I_i$'s  are  positive in the clockwise direction around the puddle.
Eqs. (\ref{Ii},\ref{Vi}) yield a simple proportionality between the Hall voltage and the tunnel current
\be
V_H \equiv \delta\mu_1-\delta\mu_3=\delta\mu_2-\delta\mu_4=\mbox{sgn}(B) {h\over e^2} k I
\label{V_H}
\ee
The relation between the {\em longitudinal} voltage drop and the tunneling current  through the barrier is
\be
V(I) \equiv \delta\mu_1-\delta\mu_2=\delta\mu_3-\delta\mu_4,
\label{VI}\ee
which is in general a non linear function. Here we assume symmetry under reversal of magnetic  field $V(I,B) = V(I,-B)$, which is expected for dissipative
current transport across a narrow channel. 

We now consider the response of a random network of puddles and tunnel barriers with two current leads at $-x$ and $+x$, and two voltage leads at   $-y$ and $+y$.
The network is described by a general  two dimensional  graph with $N_v$ 
vertices at the locations of the puddles, and  $N_b$ bonds for each  of the 
tunnel barriers (see, e.g., Fig 2).
The two dimensional layout of the puddles network ensures that bonds  do not 
cross.   

Henceforth we shall
assume that all quantum interference effects take place within the tunnel barrier lengthscales $L_{ij}$
beyond which dissipation due to low lying edge excitations destroys coherence between tunneling events. 
Thus,
the response of the puddles network is  given by  classical Kirchoff's laws. 
First, current conservation at each vertex (puddle) $i$ is given by 
\be
\sum_{j\in \{ (ij)\}} I_{ij}=0,~~~~~i=1,\ldots N_v
\label{K-Nv}
\ee
where $ \{(ij)\}$ denote the set of bonds emenating from $i$. Second, 
the sum of voltage
differences around each plaquette $p$ is given by
\be
\sum_{(ij)\in p} V_{(ij)}(I_{ij})=0,~~~~~p=1,\ldots N_p
\label{K-Np}
\ee
where $V_{(ij)}(I_{ij})$ is the non linear function of Eq. (\ref{VI}).
A total current $I$ is forced through the network through a lead coming from  $-x$ and leaving
toward $+x$. There are no currents flowing through external leads in the  
$\pm y$ directions.  It is easy to prove the following:
\newline
{\em Lemma:  The currents $I_{ij}$ in the network are completely determined by $I$.}
\newline
The proof uses Euler's theorem for two dimensional graphs \cite{EulerT}
\be
N_v + N_p -N_b =1
\ee
Thus the number of Kirchoff's equations (\ref{K-Nv}, \ref{K-Np}) is  $N_v+N_p$
which exceeds the number of unknown currents $N_b$ by one. The additional equation determines that the current flowing out of 
the $+x$ lead must be, of course,  $I$.  Q.E.D. 

As shown above, the Hall relations (\ref{V_H}) have no effect on the currents $I_{ij}$.
The total transverse voltage $V_y$ is given by choosing any path of bonds 
${\cal C}$ which connects  the $-y$ lead to the $+y$ lead (see Fig. 2)
and summing the voltages
\be
V_y= \sum_{i\in{\cal C}} V_{i,i+1}(I_{i,i+1},|B|) +\mbox{sgn}(B) k {h\over e^2}
\sum_{i\in{\cal C},(ij)'} I_{i,j}
\ee
where  $(ij)'$ denote all currents entering vertex $i$ from $-x$.  By global current conservation, the second term is proportional to the total current. Defining the Hall voltage $V_H$  to be the antisymmetric component of $V_y$ 
we thus obtain 
\be
V_H =  \mbox{sgn}(B) k {h\over e^2}I
\ee
which yields a quantized Hall resistance of  $\rho_{H}= k (h/ e^2)$
that is completely independent of $B$ and $I$.

This relation should hold as long as the network does not involve
appreciable contributions from edgestates of puddles of different $k$ values. 
The width of the QHI regime therefore depends on relative abundance of 
different density puddles, which depends in turn on the distribution of 
potential fluctuations.  As the magnetic field increases, a wide distribution 
of potential minima will create mixed phases with puddles of different  
densities. Relation  (\ref{V_H}) does not apply for tunneling between 
different $1/k$--QH liquids,
and thus the above analysis fails for the mixed phase.

\section{The Non--Linear Longitudinal Transport}
\label {sec:long}
The dissipative response in the model introduced above is
associated with the longitudinal transport through the tunnel barriers. The 
barriers connect edge--states of neighboring puddles of density 
$n=B/(k\phi_0)$, and we assume henceforth that $k$ is the same in all puddles.

A non--linear current--voltage  relation 
for a  tunnel junction between $1/k$ QH Liquids was first proposed by Wen \cite{Wen}
who mapped the fractional QH edge--states 
to chiral Luttinger Liquids. For small currents, the relation is  a power--law\cite{Wen,KF} 
\beq
I\sim \mbox{sgn}(V) V^{2g-1}\; ,
\label{ivg}
\eeq
where $g=k$ is the Luttinger liquid interaction parameter (and is equal to unity for
the integer Hall liquid).

Renn and Arovas (RA) \cite{RA} have extended Wen's result to  long tunnel 
barriers following Giamarchi and Schultz' renormalization group equations 
for  disordered Luttinger liquids\cite{GS}. They consider  the ``disordered 
antiwire'' geometry, i.e. a barrier of length with $n_t$ tunnel couplings 
of average magnitude $t$.  In the  small current limit they   obtained that 
$g$ gets renormalized $g\to \tg > k$, and the longitudinal  response is
\be
V_{RA}(I) \simeq  V_0 \mbox{sgn}(I) \left({|I|\over I_0D} \right)^{1/(2\tg-1)} ,
\label{VI-RA}
\ee
where
\bea
V_0&=& {\hbar v\over el},\nonumber\\
I_0&=& {ev\over 2\pi l k},\nonumber\\
D&\simeq& {n_t |t^2| \over 2\pi v^2 \hbar^2 }.
\label{i0-v0}
\eea
Here $v$ is the edge 
state velocity, and $l=\sqrt{\hbar c/eB}$ is the  magnetic length.

Here we consider a network of RA junctions, and assume that the
dephasing time is short enough that the tunneling events through consecutive
junctions in the network are incoherent (coherent backscattering effects are
included in RA's calculation of the single junction). Our model consists of
a random network of 
classical non--linear resistors, each characterized by a power $\alpha_n$
and a conductance prefactor $D_n$
\beq
{V_n\over V_0}=\mbox{sgn}(I)\left({|I_n|\over I_0D_n}\right)^{\alpha_n}\; ;
\label{ivan}
\eeq
By (\ref{i0-v0}), we assume that $V_0$ and $I_0$  are weakly dependent on
the barrier height fluctuations and magnetic field, compared e.g. to $D$. 
Thus  for simplicity they are taken to be uniform in the entire
network. The network of junctions with $D_n\leq 1$ is assumed to percolate 
through the sample. Thus we can choose $D_n,\alpha_n$  to be
random variables whose distribution is bounded by
\bea
0 &\leq  D_n \leq&1\nonumber\\
(2k-1) &\geq \alpha_n \geq &  1/(2k-1)
\eea
In appendix \ref{app:B} we estimate the magnetic field dependence of the average
conductance prefactor to be
\be
{\bar D}(B)\propto \exp\left(  - k n_p \left(B-B_c \over 2B_c  \right)^2  \right)
\label{barD}
\ee
where  $n_p$ is the typical number of electrons in a puddle, and $B_c$ is the
magnetic field at which the puddles percolate through the sample.
The average power law is  estimated using RA's renormalization group equations.
We find that (see App. \ref{app:B}), in the limit of small $D$ 
\be
{\bar\alpha}(B)\sim {1\over 2k-1}+{ k-3/2  \over  2k-1}{\bar D}(B)\; ,
\label{baralpha}
\ee
and for $ {\bar D}\rightarrow D_c=(2\ln 2-1)$ 
\bea
{\bar\alpha}\simeq {1\over 2+3\sqrt{D_c-\langle D\rangle}}\; .
\label{barAlpha}
\eea

We have solved  Eqs. (\ref{K-Nv}) and (\ref{K-Np}) numerically, using an IMSL Levenberg-Marquardt algorithm, for  square lattices of sizes upto
$5\times 5$. The distributions of $(D_n,\alpha_n)$ were taken to be 
\bea
P_1(D)= \Theta(D)-\Theta(D-1)\nonumber\\
P_2(\alpha)={1\over\sqrt{2\pi}\sigma}\exp\left\{{(\alpha-{\bar\alpha})^2\over
2\sigma^2}\right\}\; ,
\label{Palpha}
\eea 
We take the variance $\sigma$, according
to our estimate in appendix \ref{app:B}  to be 5 to 10 times smaller
than the mean ${\bar\alpha}$.  The numerical results in the regime 
$I_0/10 < I < I_0$, averaging over 5 realizations of disorder,  
can be summarized by the averaged network's $I-V$ response:  
\be
{ \log V \over \log I } \equiv \alpha_{eff} = {\bar\alpha} \pm \epsilon 
\label{aeff}
\ee
where $\epsilon\sim 10^{-2}-10^{-3}$. 
That is to say,  in the 
moderate current regime {\em the total voltage-current relation is given 
quite well by the  average power law}. In the extremely small current limit, 
one expects (\ref{aeff}) to break down since due to the power law resistors, 
the currents choose to flow through percolating networks of highest power 
laws. In this regime the numerical algorithm also fails to converge properly.  

In order to better estimate the corrections to the
pure power law, we  examine a toy model dubbed
the Parallel--Series (PS) network.  This model comprises of a random  
combination of serial and parallel connections of elements $P$ and $S$, where
$P$ is composed of $N_p$ resistors in parallel, and $S$ is its dual -- a 
linear chain of $N_s$ resistors in series (see Fig. 3).  
The $S$ and $P$ components can be created from
an ordinary two dimensional network by a three peaked  distribution of $D_n$'s 
(shorts, disconnections and resistors of $D_n=1$).  This model is
symmetric on average with respect to exchange of the $x$ and $y$ directions, and hence is an adequate description of macroscopically isotropic samples.

In Appendix (\ref{app:a}) we show that for currents which obey $\sigma^2\ln (I/I_0)\ll{\bar\alpha}$,
\beq
{V\over V_0}=\left({I\over I_0}\right)^{\alpha_{eff}},\quad \alpha_{eff}=
{\bar\alpha}+\left({\sigma^2\ln (I/I_0)(1-1/{\bar\alpha})\over 4}\right)\; .
\label{IVfinal}
\eeq
The deviation of $\alpha_{eff}$ from ${\bar\alpha}$ is
positive for ${\bar\alpha}<1$. This indicates that in the `insulating'  regime, 
although serial and 
parallel connections are equally represented, parallel configurations dominate at low
currents.
The situation is reversed in the QH liquid side of the transition, where 
${\bar\alpha}>1$, while at the critical filling fraction (where ${\bar\alpha}=1$)
the $S$ and $P$ elements balance each other and $\alpha_{eff}={\bar\alpha}$.
We note that under a duality transformation, which exchanges each resistor in
the network by a perpendicular resistor with $(V_n/V_0)$ and $(I_n/I_0)$ 
interchanged, the $I-V$ characteristic of the whole network is inverted: 
${\bar\alpha}\rightarrow 1/{\bar\alpha}$, $\sigma\rightarrow \sigma/{\bar\alpha}^2$,
and consequently $\alpha_{eff}\rightarrow 1/\alpha_{eff}$
which is consistent with our requirement of macroscopic  isotropy.

In comparing the results of the PS model to the square lattice simulations 
we find that the correction to a pure power law in the numerical results, 
is  smaller by at least a factor of 10 than the results of the PS model (\ref{IVfinal}). 
We suggest that the difference arises due to the fact that the PS model
assumes greater inhomogeneity  in $D_n$ as mentioned before. 
Eq.~(\ref{IVfinal}) can therefore be regarded as an estimate of the upper 
limit on the discrepancy between the macroscopic $\alpha_{eff}$ and 
${\bar\alpha}$ at moderate currents. The principal conclusion to be taken away from this calculation 
is that due to the self averaging property, the macroscopic
$I-V$   is directly related to the physics of the single junction, 
and the  non linear tunneling response between fractional quantum Hall  edge-states.

\section{Summary and Final Remarks}
\label{sec:summary}
As demonstrated in the previous sections, the QHI phase observed in proximity
to a QH liquid can be modelled by  a network of  puddles. Although similar in 
spirit to the semiclassical percolation description of Ref. \cite{CC},  
it naturally incorporates the electron interaction effects under the same 
assumption: smoothly varying potentials relative to the magnetic length $l$. 
The most important feature of this
model is that, in contrast with models based on single--electron hopping, 
it yields a 
quantized Hall resistance. The quantization is not affected by non--linearity
of the dissipative part of the response. The latter is studied for $1/k$--QHI
with $k>1$, yielding a power--law behavior of the longitudinal $I-V$ curve
which is closely determined by the behavior of an average single junction 
between adjacent puddles.  
Deviations from a pure power--law are at most of
order $\sigma^2\ln (I)$ (where $\sigma$ is the variance of the power 
distribution), estimated in appendix \ref{app:B} to be typically small. 

The magnetic field dependence of the average  tunneling rate
is gaussian as shown in Eq. (\ref{barD}), with a width defined by the inverse number of electrons in a typical puddle. Thus, smaller puddle sizes allow a larger regime of the Quantized Hall Insulator phase. However, 
if these incompressible  puddles  are too small, it means that our assumption of slowly
varying potential becomes invalid.

We note that the integer QH case of $k=1$ implies all $\alpha_n=1$ throughout the network. That is to say, that the puddles model reduces naturally to a random Ohmic resistors network with conductances proportional to Eq.( \ref{barD}). Interference effects between junctions \cite{CC} are ignored here, since we assume an inelastic scattering length of the order of inter-junction separation, 
an assumption that breaks down at low enough $T$.

Our analysis so far has concentrated on the non linear transport of tunnel junctions,
applicable for large enough bias and low $T$. 
At finite $T$, transport in the junctions -- and 
hence through the entire network -- crosses over to linear response at 
sufficiently low bias  $ khi/e < k_BT $, where $i=I/N_y$ is the average
current through  single--junction and $N_y$ is the typical number of junctions
across the sample.
The linear conductivity is
then predicted to vary as a power--law of temperature \cite{RA}, i.e. 
$V\propto IT^{1-1/{\bar\alpha}}$. A temperature--dependence 
measurement of the resistance in the Ohmic regime can thus provide a 
further test of our model. In addition, for a given $T$ the crossover from a
linear to non--linear $I(V)$ can provide an estimate of $N_y$.

One of the most interesting implications of our suggested puddles--network 
model is that the insulating phase, surrounding the fundamental QH liquids
in the phase diagram of Ref. \cite{KLZ}, is not a homogeneous phase. 
Restricted regions in the phase diagram which are in proximity to specific
$1/k$--QH liquids are dominated by weakly coupled puddles of the corresponding
liquid. It is therefore implied, that a measurement of $\rho_{xy}$ as a 
function of magnetic field at moderate disorder may exhibit plateaux at odd
integer multiples of $h/e^2$ - even though the longitudinal transport 
indicates an insulating--like behavior. The width of the plateaux is expected 
however, to depend on details of the disorder potential in the sample. The
width of the ``mixed--phases'', where $\rho_{xy}$ rises with magnetic field 
between consecutive plateaux, is 
{\it not} expected to vanish for $T\rightarrow 0$ as in the QH liquid regime.

Finally, we would like to comment on an open problem with regards to comparison
of this theory with experimental results of Ref. \cite{SSS}. The experiment has indicated a duality symmetry
between  $I-V$ curves at 
opposite sides of the $1/3$--QH liquid--to--insulator transition. This 
phenomenon was interpreted in terms of charge--vortex duality, 
or equivalently as particle--hole symmetry\cite{SSS}.  In the puddles network 
model, such duality would be observed if at  $B>B_c$ each tunnel barrier with 
response  $I=F(V)$ is related to  a narrow channel formed at $B'<B_c$,  such 
that $I' =F^{-1}(V')$. However, recent theories for a single scatterer 
in a narrow channel \cite{Wen,KF} do not
yield this relation. The multiple tunneling case \cite{RA} has only  treated 
electron tunneling in the large barrier limit of $B>B_c$.
Resolution of this point is left to further research.

\acknowledgements
We thank  D. Shahar for discussions that motivated this work, and 
acknowledge useful conversations with D. Arovas, Y. Avron, D. Bergman, 
Y. Gefen, C. Henley and U. Sivan. This work was partly supported by the
Technion -- Haifa University Collaborative Research Foundation, the Fund for Promotion of Research at Technion, and a grant from the Israeli Academy of Sciences.

\appendix
\section{Dependence of Distributions on Magnetic field}
\label{app:B}
To facilitate a comparison with experiment, we must somehow relate the
average and mean square deviations 
of $D_n$ and $\alpha_{n}$ in Eqs. (\ref{Palpha})
to the
external magnetic field $B$. Here we make substantial use of the results of Renn and Arovas \cite{RA}, which allow us to express $D$ and $\alpha$ in terms of the semiclassical tunneling
probability at the junction.  
The renormalization group equations of \cite{RA,GS} connect $\alpha$ to $D$
as follows. In the insulating limit
\beq
D_n\rightarrow 0:~~~~~~~~\alpha_{n}={1\over 2g_n-1}\simeq {1\over 2k-1}+{(k-3/2)D_n\over 2k-1}\; ;
\label{Tn-0}
\eeq
in the regime $D_n\simeq D_c\equiv (2\ln 2-1)$, we get
\beq
\alpha_{n}\simeq{1\over 2+3\sqrt{D_c-D_n}}\; .
\label{Tn-Tc}
\eeq    
Eqs.~(\ref{Tn-0}), ~(\ref{Tn-Tc}) relate $\alpha_{n}$ to  $D_n$ in the range
$1/(2k-1)\leq \alpha_{n}\leq 1/2$; the analysis of \cite{RA,GS} is not 
applicable closer to the QH liquid/insulator transition, where $1/2<\alpha<1$.
Note that the effect of increasing tunneling rate is to interpolate between the
limits $\alpha_{n}=1/(2k-1)$ and $\alpha_{n}=1/2$.

We assume that the potential fluctuations are bounded, and have a characteristic length scale of fluctuations $l_V$.  This lengthscale also represents the typical linear size of the puddles, which will turn out to be an important parameter in the following discussion.

Since the puddles are incompressible, a change in the magnetic field $\delta B$ near the percolation field $B_c$ will shrink the puddles by a linear distance $\delta l$ which is related to $\delta B$ by
\beq
{\delta l\over l_V}={\delta B\over 2B_c}
\label{lp-B}
\eeq

The  tunneling rate of an electron in the lowest Landau level through  a quadratic potential barrier $V(x,y)= {1\over 2} V'' (-x^2 + y^2)$ is solved by mapping the problem to a one dimensional hamiltonian given by
\be
H=   {1\over 2m} p^2 - {1\over 2} V''  x^2 
\ee
where the `tunneling mass' is $m= \hbar^2 /V''l^4   $ ($l$ being the magnetic
length). Using the WKB expression for tunneling at energy $-V_B$,   
\bea
D&=&  D_0 \exp\left( -{2\pi\over \hbar} V_B 
\sqrt{m\over V''}\right)\nonumber\\
&=& D_0 \exp\left( -\pi 
\left(\delta l \over l \right)^2 \right)\nonumber\\
&=& D_0 \exp\left( - \pi 
\left(\delta B \over 2B_c  \right)^2\left( l_V  / l \right)^2 \right) 
\eea
The factor $x=\left( l_V  / l \right)^2$ is roughly $k$ times the number of 
electrons in the puddle, and it determines the field dependence of the 
tunneling rate near percolation. In a specific junction in the network, 
it is given by a random variable  $x_n$ with 
average $\langle x\rangle$ and variance $\sigma_x$ such that
\bea
D_n & \sim&  D_0 \exp\left( -x_n  \left(\delta B \over 2B_c  \right)^2
\right) \nonumber\\
\langle D\rangle &=& D_0\exp\{-\langle x\rangle (\delta B/ B_c)^2\}\nonumber\\ 
\quad \sigma_D & \simeq &\left({\delta B\over B_c}\right)^2\sigma_x\langle D\rangle 
\label{Rav}
\eea
Employing Eqs. (\ref{Tn-0}) and (\ref{Tn-Tc}),
we find that ${\bar\alpha}$ and $\sigma$ of section \ref{sec:long} are given by
\bea
{\bar\alpha}\simeq {1\over 2k-1}+{(k-3/2)\langle D\rangle\over  2k-1}
\; ,\nonumber\\
\sigma\simeq {(k-3/2)\sigma_D\over  2k-1}
\label{alphaTn-0}
\eea
for $\langle D\rangle\rightarrow 0$, and
\bea
{\bar\alpha}\simeq {1\over 2+3\sqrt{D_c-\langle D\rangle}}
\; ,\nonumber\\
\sigma\simeq \left({3\over 2{\bar\alpha}^2}
{1\over\sqrt{D_c-\langle D\rangle}}\right)\sigma_D
\label{alphaTn-Tc}
\eea
for $\langle D\rangle\rightarrow D_c$. Substituting, e.g., $k=3$ and 
$\delta B/B_c<10^{-1}$,
we find that $\sigma$ is typically 10 times smaller than ${\bar\alpha}$, which is
the values we have used in the numerical simulations of the square lattice network.

\section{Derivation of $V(I)$ in the PS Model}
\label{app:a}
Consider a system of $S$ and $P$ elements, which are serial and parallel
connections of power--law resistors, respectively.
We first derive the local current--voltage response of elements $P$
and $S$ separately. In $P$, the average current per parallel unit is related
to the voltage $V_p$ by
\beq
I_p=\langle{1\over N_p}\sum_{n=1}^{N_p}I_n\rangle=
I_0\langle{1\over N_p}\sum_{n=1}^{N_p}
\left({V_{p}\over V_0}\right)^{1/\alpha_{n}}\rangle\; ,
\label{AvIp}
\eeq 
where the angular brackets denote averaging over the distribution 
$P_2(\alpha)$ (Eq.~(\ref{Palpha})), which yields
\beq
{I_p\over I_0}={1\over\sqrt{2\pi}\sigma}\int_{-\infty}^{\infty}d\alpha 
\exp\left\{{(\alpha-{\bar\alpha})^2\over 2\sigma^2}\right\}
\left({V_{p}\over V_0}\right)^{1/\alpha}\; .
\label{Ipint}
\eeq
In the saddle point approximation (valid for 
$\sigma^2\ln (I_p/I_0)\ll{\bar\alpha}$),
\beq
{V_{p}\over V_0}=\left({I_p\over I_0}\right)^{\alpha_p}\; ,\quad
\quad\alpha_p={\bar\alpha}-{\sigma^2 \ln (I_p/I_0)\over 2{\bar\alpha}}\; .
\label{VpIp}
\eeq
Note that the approximation breaks down in the limit of very small currents.
Eq.~(\ref{VpIp}) implies that the contribution of a purely parallel 
configuration to the network over--weighs the significance of better 
conducting channels. This produces a positive shift of the power--law which
is enhanced at small currents. The response of a single $S$-type element 
indicates an opposite trend: similarly to  Eq.~(\ref{VpIp}), the average
voltage $V_s$ (per serial unit) is related to the current $I_s$ by
\beq
{V_s\over V_0}=\langle{1\over N_s}\sum_{n=1}^{N_s}{V_n\over V_0}\rangle=
{1\over\sqrt{2\pi}\sigma}\int_{-\infty}^{\infty}d\alpha 
\exp\left\{{(\alpha-{\bar\alpha})^2\over 2\sigma^2}\right\}
\left({I_{s}\over I_0}\right)^{\alpha}\; ,
\label{Vsint}
\eeq
and hence
\beq
{V_{s}\over V_0}=\left({I_s\over I_0}\right)^{\alpha_s}\; ,\quad 
\quad\alpha_s={\bar\alpha}+{\sigma^2 \ln (I_s/I_0)\over 2}\; .
\label{VsIs}
\eeq
The negative shift of the effective power reflects the over--emphasis of the
larger resistors in the chain, which is particularly pronounced at small
currents.

We next consider the overall response of $N$ serially connected elements, of 
type $P$ and $S$ alternately. Denoting $I=I_p=I_s$, we get
\beq
V=\langle{1\over N}\sum_{n=1}^{N}V_n\rangle=
{V_0\over 2}\left(\left({I\over I_0}\right)^{\alpha_p}+
\left({I\over I_0}\right)^{\alpha_s}\right)\; ,
\label{AvVps}
\eeq
and thus, for $\sigma^2\ln (I/I_0)\ll{\bar\alpha}$,
\bea
{V\over V_0}&=&\left({I\over I_0}\right)^{\alpha_{eff}}\; ,\nonumber\\
\alpha_{eff}&=&{(\alpha_p+\alpha_s)\over 2}={\bar\alpha}+
\left({\sigma^2\ln (I/I_0)(1-1/{\bar\alpha})\over 4}\right)\; .
\label{IVfinalApp}
\eea
It is straight--forward to show that a parallel connection of alternating 
$P$ and $S$ type elements yields the same effective power--law. We therefore
conclude that {\it any} configuration which involves serial and parallel 
connections of evenly distributed $P$ and $S$ type elements will have a 
current--voltage characteristic given by Eq.~(\ref{IVfinalApp}).

\begin{figure}[htb]
\centerline{\psfig{figure=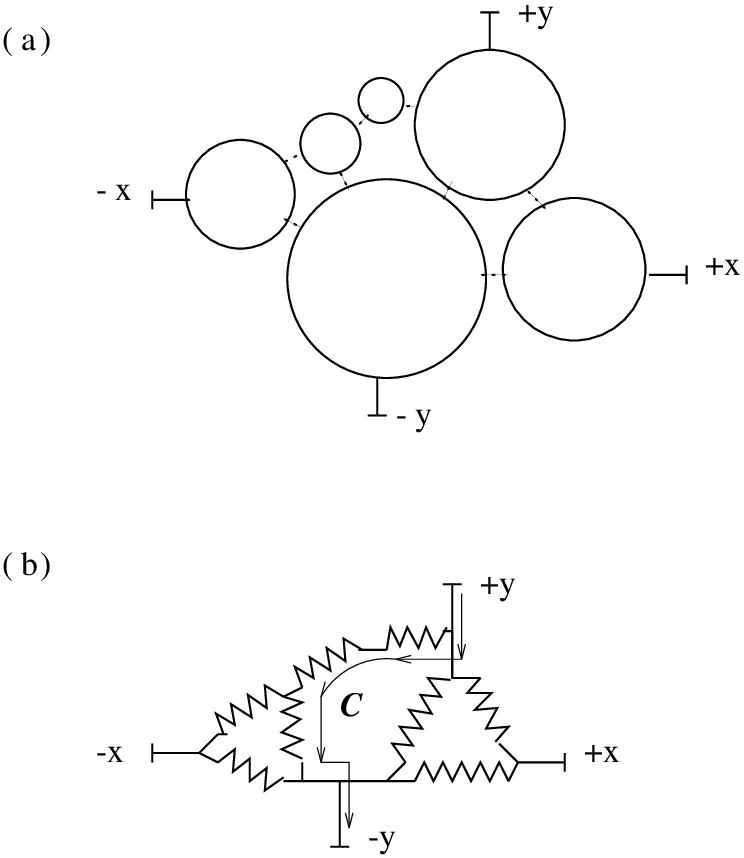,width=3.5in}}
\vspace{0.5in}
\caption[]
{
\label{fig:fig2}
(a) A typical puddles--network (with $N_v=6$, $N_b=8$). (b) The corresponding 
equivalent circuit. The path ${\cal C}$ is denoted by arrows. 
}
\end{figure}
\begin{figure}[htb]
\centerline{\psfig{figure=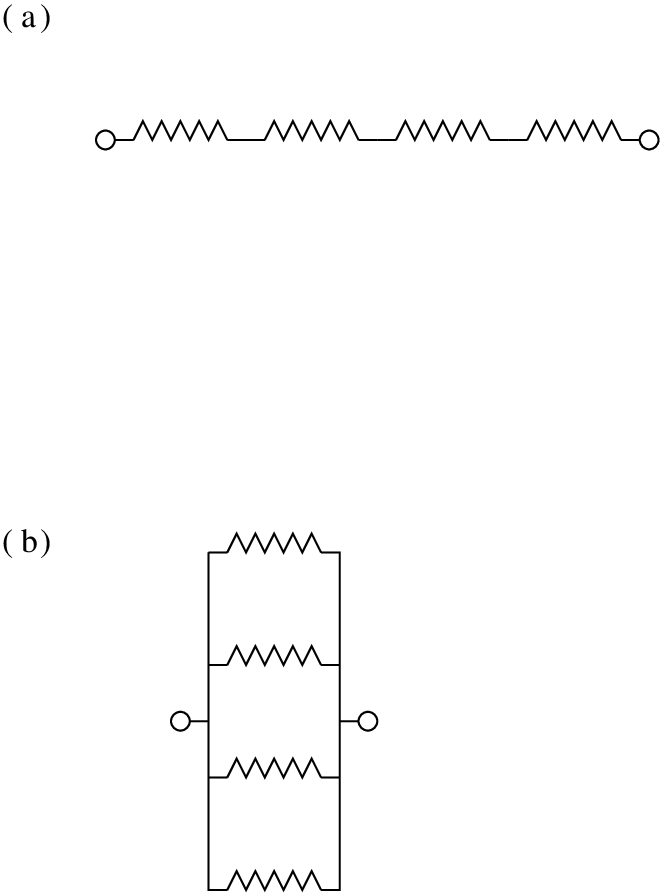,width=4in}}
\vspace{0.5in}
\caption[]
{
\label{fig:fig3}
(a) An $S$--type element ($N_s=4$). (b) A $P$--type element ($N_p=4$).
}
\end{figure}


\begin{references}
\bibitem{Fuku}
H. Fukuyama, J. Phys. Soc. Jpn. {\bf 49}, 644 (1980); B. Altshuler, 
D. Khmelnitskii, A. Larkin and P. A. Lee, Phys. Rev. B {\bf 22}, 5142 (1980).
\bibitem{KLZlett}
S. C. Zhang, S. Kivelson and D. H. Lee, Phys. Rev. Lett. {\bf 69}, 1252 (1992).
\bibitem{Joe}
Y. Imry, Phys. Rev. Lett. {\bf 71}, 1868 (1993).
\bibitem{Ora}
O. Entin-Wohlman, A. G. Aronov, Y. Levinson and Y. Imry, Phys. Rev. Lett. 
{\bf 75}, 4094 (1995).
\bibitem{Hop}
P. Hopkins {\it et al.}, Phys. Rev. B {\bf 39}, 12708 (1989).
\bibitem{Vladimir}
V. J. Goldman, M. Shayegan and D. C. Tsui, Phys. Rev. Lett. {\bf 61}, 881 
(1988); V. J. Goldman, J. K. Wang, B. Su and M. Shayegan, Phys. Rev. Lett.
{\bf 70}, 647 (1993); R. L. Willett, H. L. Stormer, D. C. Tsui, L. N. Pfeiffer,
K. W. West and K. W. Baldwin, Phys. Rev. B {\bf 38}, 7881 (1988).
\bibitem{SSS}
D. Shahar, D. C. Tsui, M. Shayegan, E. Shimshoni and S. L. Sondhi, 
{\it preprint} (1995).
\bibitem{KLZ}
S. Kivelson, D. H. Lee and S. C. Zhang, Phys. Rev. B {\bf 46}, 2223 (1992).
\bibitem{CC}
J. T. Chalker and P. D. Coddington, J. Phys. C {\bf 21}, 2665 (1988).
\bibitem{ruzin}
Quantization of $\rho_{xy}$ has been demonstrated for the critical 
regime: see A. M. Dykhne and I. M. Ruzin, Phys. Rev. B {\bf 50}, 2369 (1994);
I. M. Ruzin and S. Feng, Phys. Rev. Lett. {\bf 74}, 154 (1995).
\bibitem{RA} S.R. Renn and D.P. Arovas, Phys. Rev. {\bf B51}, 16832 (1995).
\bibitem{Beenakker} C.W.J. Beenakker and H. van Hoiuten, {\it Solid State Physics: Advances in Research and Applications}, Ed. H. Ehrenreich and D. Turnbull (
Academic, San Diego, 1991), Vol 44, pp.207, 208.
\bibitem{EulerT}
This relation is more commonly known as $V+F-E=\chi$, where $V,F,E,\chi$ are 
the vertices, faces, edges and Euler characteristics of general
polygons in three dimensions.
\bibitem{Wen} X.-G. Wen, Int. Journ. Mod. Physics B, {\bf 6}, 1711 (1992).
\bibitem{KF}
C. L. Kane and M. P. A. Fisher, Phys. Rev. B {\bf 46}, 15233 (1992).
\bibitem{GS} T. Giamarchi and H.J. Schultz, Phys. Rev. {\bf B37}, 325  (1988). 

\end{references}
\end{document}